**Single spin universal Boolean logic**

H. Agarwal[1], S. Pramanik and S. Bandyopadhyay

Department of Electrical and Computer Engineering

Virginia Commonwealth University

Richmond, VA 23284, USA

ABSTRACT

Recent advances in manipulating single electron spins in quantum dots have brought us close to the realization of classical logic gates based on representing binary bits in spin polarizations of single electrons. Here, we show that a linear array of three quantum dots, each containing a single spin polarized electron, and with nearest neighbor exchange coupling, acts as the universal NAND gate. The energy dissipated during switching this gate is the Landauer-Shannon limit of $kTln(1/p_i)$ [$T$ = ambient temperature and $p_i$ = intrinsic gate error probability]. With present day technology, $p_i = 10^{-9}$ is achievable above 1 K temperature. Even with this small intrinsic error probability, the energy dissipated during switching the NAND gate is only ~ 21 $kT$, while today's nanoscale transistors dissipate about 40,000 – 50,000 $kT$ when they switch.

Corresponding author S. Bandyopadhyay, e-mail: sbandy@vcu.edu

---

[1] Undergraduate summer intern visiting from Banaras Hindu University Institute of Technology, Varanasi, India.

1. Introduction

The primary threat to continued downscaling of electronic devices envisioned in Moore's law[1] is excessive energy dissipation that takes place when a device switches between logic levels. If electronic devices are shrunk relentlessly without concomitantly reducing energy dissipation, thermal management on a chip will ultimately fail resulting in chip meltdown. Conventional devices have a fundamental drawback in this regard since they typically encode digital information using charge (or voltage/current levels determined by charge). Charge is a scalar quantity that has only a magnitude. Therefore, binary logic bits 0 and 1 must be demarcated by a difference in the magnitude of the charge representing the bit. Switching between bits would mandate changing this magnitude, which invariably involves current flow and associated power dissipation of $I^2R$ (*I*=current and *R*=resistance in the path of the current). There is no way to avoid this dissipation.

Spin, on the other hand, is a pseudo vector that has both a magnitude and a polarization. The polarization can be made *bistable* by placing the electron in a dc magnetic field, so that only two polarizations – parallel and anti-parallel to the field - are stable. They can encode the bits 0 and 1. Switching between them requires simply flipping the spin without physically moving the charge in space and causing a current flow. This can reduce energy dissipation significantly.

In this paper, we first show rigorously how a universal Boolean logic gate (the NAND gate) can be realized based on this idea. Although the basic idea was proposed many years ago[2], this is the first rigorous quantum mechanical calculation establishing the gate's *truth table*. Next, we present an estimate of the energy dissipated during switching this gate. The energy dissipated is found to be the minimum allowed by the laws of thermodynamics, namely the Landauer-Shannon limit[3].

## 2. A single spin Boolean NAND gate.

Consider a linear array of three single electron containing quantum dots shown in Fig. 1. The quantum mechanical wavefunctions of electrons in nearest neighbor dots overlap in space causing exchange coupling between them. A weak global magnetic field makes the spin polarization in each dot bistable, because the polarization can be either parallel or anti-parallel to the global field. These two stable polarizations encode the classical binary bits 1 and 0, respectively.

The two peripheral dots **A** and **C** host the two input bits and the central dot **B** hosts the output bit. Input data are provided by orienting the spins in **A** and **C** in the desired directions (parallel or anti-parallel to the global magnetic field) with local magnetic fields generated by local inductors, as in magnetic random access memory (MRAM) chips. These inductors are wrapped around the input dots and might be realized with carbon nanotubes. We will show that when the system is in the ground state, the output spin polarization in dot **B** always conforms to the NAND function of the input spins in dots **A** and according to the truth table of a NAND gate:

**Table I: Truth Table of the NAND gate**

| Input 1 (A) | Input 2 (C) | Output (B) |
|---|---|---|
| 1 | 1 | 0 |
| 0 | 0 | 1 |
| 0 | 1 | 1 |
| 1 | 0 | 1 |

The output can be read with a variety of techniques that are capable of single spin detection[4-6].

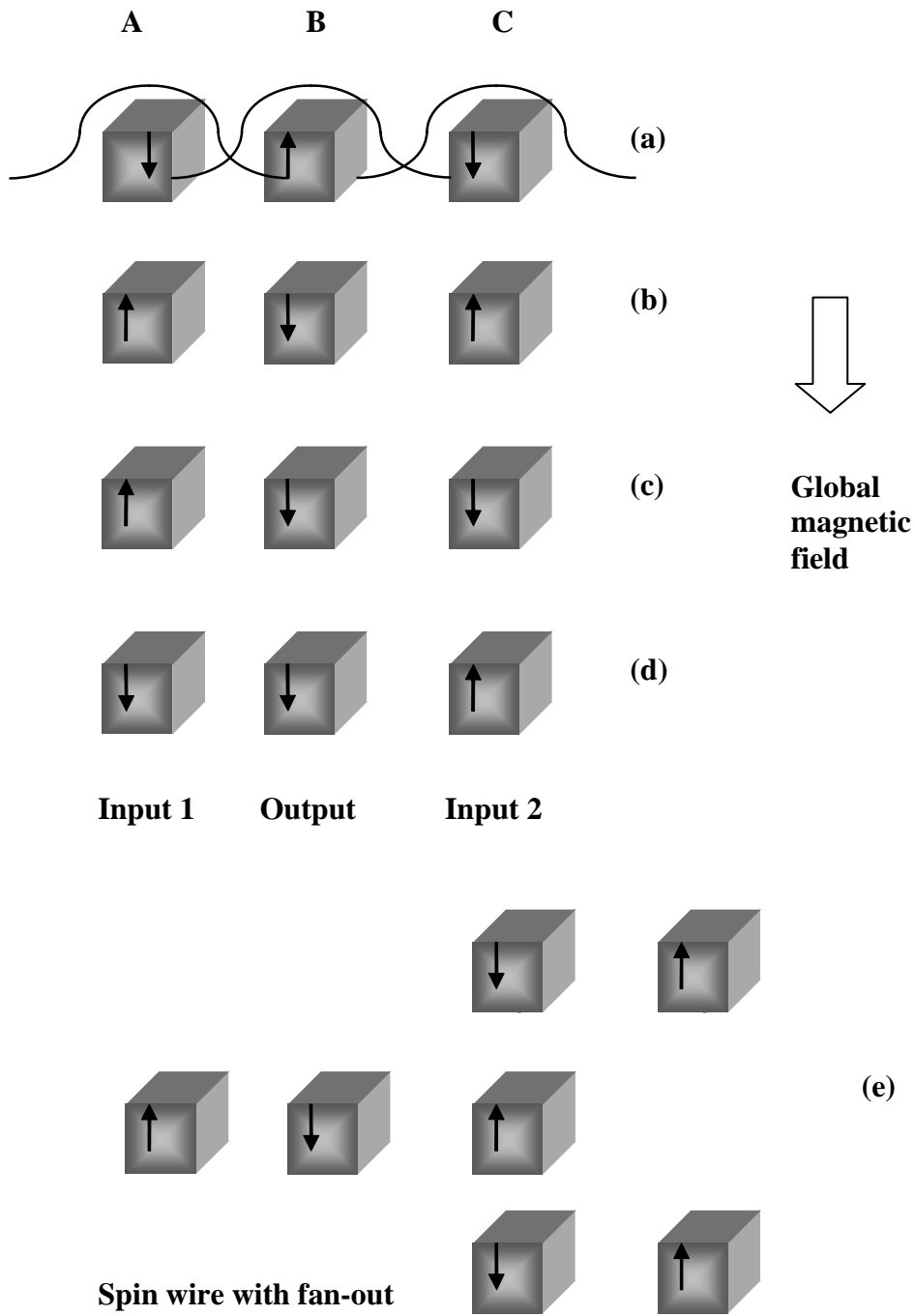

**Fig. 1: A single spin NAND gate**: *An array of 3 spin polarized single electrons, each housed in a quantum dot, realizes the NAND gate when the entire array is placed in a static magnetic field and allowed to relax to the thermodynamic ground state. Wavefunctions of nearest neighbor electrons overlap*

*(see row (a)) resulting in nearest neighbor exchange coupling. The two peripheral spins **A** and **C** are input bits and the central spin **B** is the output bit. Downspin polarization (parallel to the static magnetic field) corresponds to logic bit 1 and upspin (anti-parallel to the magnetic field) to logic bit 0. When the spin polarizations in dots **A** and **C** are aligned to conform to the desired input bits, and the system is allowed to relax to the ground state, the spin polarization in dot **B** always corresponds to the NAND function of the inputs. Shown in Figs. 1(a) – 1(d) are the spin configurations in all three dots corresponding to the four possible input combinations of a primitive NAND gate. Fig. 1(e) shows a "spin wire" [with fan out] in which the spin signal is replicated in every other dot. To transmit signal unidirectionally along a spin wire, clock pads are interposed between every dot pair. The voltages at these pads are raised sequentially using a 3-phase clock to ferry the spin signal unidirectionally. The clock pads are not shown explicitly in Fig. 1(e) for the sake of clarity.*

### 3. Theory

If the charging energy (intradot Coulomb repulsion) within each dot is sufficiently strong, then at half-filling (1 electron per dot), then the Hubbard Hamiltonian representing the 3-spin array in Fig. 1 will reduce to the simple Heisenberg Hamiltonian[7, 8].

$$H_{Heisenberg} = \sum_{<ij>} J_{ij}^{\Box} \sigma_{zi}\sigma_{zj} + \sum_{<ij>} J_{ij}^{\perp} \left(\sigma_{xi}\sigma_{xj} + \sigma_{yi}\sigma_{yj}\right) + \sum_{\text{input dots}} \sigma_{zi} h_{zi}^{inputs} + \sum_i \sigma_{zi} h_{zi}^{global}$$

where the $\sigma$-s are Pauli spin matrices. We assume that the lowest orbital states in the quantum dots are occupied. If excitation to the higher orbital states is not accompanied by spin flip, then the higher states do not matter in the ensuing analysis, since logic bits are encoded in the spin and not the orbital quantum number. Even if the rate of transition between orbital states is moderately high[9], the rate of spin flip is still very small[10], indicating that most excitations are not accompanied by spin flips. Consequently, the excited states are not important in our context and we do not have to consider them. We adopt the convention that the direction of the local and the global magnetic fields is the *z*-direction. The last two terms in the above equation account for the Zeeman energies associated with these fields. The first two terms account for exchange interaction between nearest neighbors (the angular brackets denote summation over nearest neighbors). We will assume the isotropic case when $J_{ij}^{\perp} = J_{ij}^{\Box} = J$, where *J* is the exchange energy, which is non-zero if the wavefunctions in dots *i* and *j* overlap in space.

The spins in the quantum dots are polarized in either the +z or –z direction (because of the global and local magnetic fields), which we designate as "upspin" (↑) and "downspin" (↓) states, respectively. We will assume that the upspin state (aligned anti-parallel to the global magnetic field) encodes bit 0 and downspin state (parallel to the global field) encodes bit 1.

The 3-spin basis states representing the spin configurations in the 3-dot array are $|\downarrow\downarrow\downarrow\rangle, |\downarrow\downarrow\uparrow\rangle, |\downarrow\uparrow\downarrow\rangle, |\downarrow\uparrow\uparrow\rangle, |\uparrow\downarrow\downarrow\rangle, |\uparrow\downarrow\uparrow\rangle, |\uparrow\uparrow\downarrow\rangle, |\uparrow\uparrow\uparrow\rangle$ where the first entry is the spin polarization in dot **A**, the second in dot **B** and the third in dot **C**. These eight basis functions form a complete orthonormal set. The matrix elements $<\phi_m | H_{Heisenberg} | \phi_n>$ are given in the matrix below, where the $\phi_{m,n}$ are the 3-electron basis states enumerated above.

$$\begin{pmatrix} 2J - h_A - h_C - 3Z & 0 & 0 & 0 & 0 & 0 & 0 & 0 \\ 0 & -h_A + h_C - Z & 2J & 0 & 0 & 0 & 0 & 0 \\ 0 & 2J & -2J - h_A - h_C - Z & 0 & 2J & 0 & 0 & 0 \\ 0 & 0 & 0 & -h_A + h_C + Z & 0 & 2J & 0 & 0 \\ 0 & 0 & 2J & 0 & h_A - h_C - Z & 0 & 0 & 0 \\ 0 & 0 & 0 & 2J & 0 & -2J + h_A + h_C + Z & 2J & 0 \\ 0 & 0 & 0 & 0 & 0 & 2J & h_A - h_C + Z & 0 \\ 0 & 0 & 0 & 0 & 0 & 0 & 0 & 2J + h_A + h_C + 3Z \end{pmatrix}$$

In the above matrix, 2Z is the Zeeman splitting energy in any dot caused by the global magnetic field, while $2h_A$ and $2h_C$ are Zeeman splitting energies in the input dots due to the local magnetic fields that write input data. If the local magnetic field is in the same direction as the global field and writes bit 1, then the corresponding *h* is positive; otherwise, it is negative. The quantity *J* is always positive (to guarantee that the singlet state composed of two coupled electrons has lower energy than the triplet state, as it should be).

In the Appendix, we tabulate the 8 eigenenergies $E_n$ ($n = 1....8$) and the corresponding eigenstates

$$\psi_n = c_1^n |\downarrow\downarrow\downarrow\rangle + c_2^n |\downarrow\downarrow\uparrow\rangle + c_3^n |\downarrow\uparrow\downarrow\rangle + c_4^n |\downarrow\uparrow\uparrow\rangle + c_5^n |\uparrow\downarrow\downarrow\rangle + c_6^n |\uparrow\downarrow\uparrow\rangle + c_7^n |\uparrow\uparrow\downarrow\rangle + c_8^n |\uparrow\uparrow\uparrow\rangle$$
$$= [c_1^n, c_2^n, c_3^n, c_4^n, c_5^n, c_6^n, c_7^n, c_8^n]$$

of the 3-dot system obtained by evaluating the eigenvalues and eigenvectors of the above 8×8 matrix. This exercise has been carried out for four cases: $h_A = \pm h$ and $h_C = \pm h$, which correspond to the four possible input combinations, and therefore the four entries in the truth table of the NAND gate.

In the Appendix, we also show that if we apply sufficiently strong local magnetic fields to orient the spins in input dots **A** and **C** to the desired directions (i.e. $|h_A|=|h_C|=h >> J, Z$), and allow the system to relax to the ground state, then the spin polarization in the output dot **B** will always represent the result of NAND Boolean logic operation on the input bits. In other words, the 3-dot system *realizes a NAND gate* whenever it is in the ground state. Ground state relaxation is the central idea in some types of artificial neural networks and similar ideas are found in other contexts[11] as well. Of course, applying local magnetic fields exclusively to specific quantum dots that host input bits is difficult and requires sophisticated shielding techniques and extreme spatial resolution. We visualize using spin polarized scanning tunneling microscope tips for this purpose, which can concentrate a magnetic field over a single quantum dot. A magnetic shield can be wrapped around each dot for further field containment. This is a difficult engineering challenge but not unachievable because of any fundamental physical laws.

Once the NAND gate is realized, we need only one other component to implement any arbitrary Boolean logic circuit. That element is a "spin wire" which will ferry spin signal unidirectionally from one gate to another. A spin wire consists of a linear array of quantum dots with clock pads between them [Fig. 1(e)]. When the clock signal at a given pad is high, the potential barrier between the two flanking dots is lowered, and their resident electrons are exchange coupled. This renders their spins anti-parallel[12]. Therefore, by sequentially clocking the barriers, we can replicate the spin bit in every other dot and transmit the spin signal along the wire unidirectionally[13]. A 3-phase clock is required for this purpose[13].

4. Gate errors

It is the natural tendency of any physical system to relax to the ground state, which is the basis of the NAND operation. However, once a system relaxes to ground state, it need not stay there forever. If it gets out of the ground state, and it does because of noise and fluctuations[14], it will produce wrong results and cause errors in computation. We will compute this error probability next.

Consider a system which is thermodynamically coupled to its environment that allows it to relax to the ground state. Once the system has attained equilibrium with the environment, the probability of finding it in any particular state is given by the Fermi-Dirac occupation probability. This probability is *not* unity for the ground state, implying that the system does not have to remain in the ground state in perpetuity. If we approximate the Fermi Dirac statistics with the Boltzmann statistics, then the ratio of the probabilities of the gate being in an excited state and the ground state is $exp[-(E_{excited} – E_{ground})/kT]$. We can call this the error probability $P_{error}$ associated with being in the excited state, since straying from the ground state causes an error in the result. The total error probability is the sum of $P_{error}$ carried out over all excited states. We will call this total error probability the *intrinsic* gate error probability $p_i$ since it accrues from the intrinsic dynamics of the gate (thermodynamics). In order to calculate $p_i$, we must first find the energy differences between the ground state and the excited states. Since the ground state and the excited state energies depend on the input bits, we have to consider all four input combinations. Referring to the energy eigenstates tabulated in the Appendix, we find:

**Case I – when inputs are [1 1]:** Here $E_1 – E_{ground} \approx 4J - 2Z$ and $E_2 – E_{ground} \approx 2h + 2Z + 2J \approx 2h$ if we take into account the fact that $h >> J, Z$. Because of the last inequality, we only have to worry about the first excited state $E_1$, since the second excited state $E_2$ is far above in energy than $E_1$.

**Case II - inputs are [0 0]:** Here, $E_1 – E_{ground} \approx 4J + 2Z$ and $E_2 – E_{ground} \approx 2h - 2Z + 2J \approx 2h$ if we again take into account the fact that $h >> J, Z$. Therefore, the same considerations as Case I apply and we only need to worry about the first excited state.

**Cases III and IV - inputs are [0 1] or [1 0]:** In these cases also, we need to worry only about the first excited states as long as $h >> J, Z$, since the second excited states are far above in energy than the first excited states. Here, $E_1 – E_{ground} \approx 2Z$.

Since the total error probability is $p_i = \sum_{\text{all excited states}} P_{error} = \sum_{m=1}^{7} e^{-(E_m - E_{ground})/kT} \approx exp[-(E_1 - E_{ground})/kT]$, we obtain $E_1 - E_{ground} = kTln(1/p_i)$. Considering the four cases above, it is obvious that we need two conditions to be fulfilled: (i) $Z = (1/2)kTln(1/p_i)$ [Case III or IV], and (ii) $4J - 2Z = 4J - kTln(1/p_i) = kTln(1/p_i)$ [Case I], which yields $J = (1/2)kTln(1/p_i)$. These conditions determine the values of $J$ and $Z$ required to restrict the intrinsic gate error probability to no more than $p_i$ at a temperature $T$.

There is actually a second source of gate errors caused by random, spontaneous spin flips that occur outside the computation sequence because of extraneous influences (e.g., spin-orbit coupling) causing spin relaxation. We call the associated gate error probability the *extrinsic* error probability $p_e$ since it accrues from extrinsic factors. It is easy to see that $p_e = 1 - exp[-T/T_1]$ where $T$ is the clock period and $T_1$ is the spontaneous spin flip time in a single quantum dot (for a single electron uncoupled with its neighbors). There are reports of $T_1 = 170$ msec – 1 sec in GaAs quantum dots[10] at low temperatures and nearly 1 second in organic nanostructures[15] at even 100 K.

We emphasize that $T_1$ is the spin relaxation time of a single electron in an isolated quantum dot that is *uncoupled* to any of its nearest neighbors. When relaxation to ground state takes place in the logic gates during computation, each electron is exchange coupled to its nearest neighbors. This relaxation is governed by the many-body relaxation of coupled spins. The single particle spin relaxation can be orders of magnitude slower than the many body spin relaxation. This is well known in the context of the transverse relaxation time $T_2$ [16]. Therefore, the relaxation to ground state can occur in a time much shorter than $T_1$. That means that the clock frequency is not limited by $1/T_1$, but can be much higher.

In ref. 13, we showed that the relaxation to ground state occurs when the clock signal is high and the nearest neighbors are coupled by exchange interaction. Therefore, the rate of relaxation to ground state is the many body rate $1/T_1^*$ which is much higher than the single particle rate $1/T_1$. When the clock signal is low, the system is in the standby state, and we would not like the spin to flip spontaneously during this time since that would cause an error. However, in the standby state, each electron is uncoupled to its neighbors and hence the spin flip rate is $1/T_1 << 1/T_1^*$.

The rate $1/T_1^*$ is obviously the upper limit on the clock frequency, since otherwise the relaxation to ground state will not be complete before the clock signal changes. The extrinsic error probability will be then limited by $p_e = 1 - exp[-T_1^*/T_1] \approx T_1^*/T_1$ if $T_1^* << T_1$.

The intrinsic and extrinsic error probabilities are not related to each other and are independent quantities. The net error probability that error correction schemes will have to contend with is the larger of $p_i$ and $p_e$. Modern error correction algorithms can handle net error probabilities as high as 3% [17].

## 5. Energy dissipation during switching

The maximum energy dissipated during switching the NAND gate is the largest energy difference between any two of the four ground states corresponding to the four input combinations shown in Figs. 1(a) – Fig. 1(d). By considering all the four ground state energies (see Appendix), the largest energy difference between any two ground states is $2Z$ (corresponding to switching between the states in Figs. 1(a) and 1(b)), which we have just shown is $kTln(1/p_i)$. Therefore, ideally, the maximum energy dissipated during switching is $kTln(1/p_i)$. This is the well-known Landauer-Shannon limit[3].

It is interesting to note that when we switch between some of the states, (e.g. between Fig. 1(b) and 1(c)), the energy dissipated is *less* than the Landauer-Shannon limit of $kTln(1/p_i)$. This happens because of *interactions* between the spins (internal feedback) which make all 3 spins act in concert as a single entity. A similar situation was addressed in ref. [18].

In the energy calculation, we purposely ignored the energy cost of generating the local magnetic fields and the energy dissipated in the clock pads. These costs can be made arbitrarily small, certainly much smaller than $kTln(1/p_i)$, by adopting adiabatic schemes[19].

6. **Temperature of operation**

The requirement $J=Z=(1/2)kTln(1/p_i)$ will also determine the maximum temperature at which we can operate the logic gates if, because of technological constraints, we are limited to specific values of $J$ or $Z$ and wish to limit $p_i$ to a specific value. With today's technology, the exchange coupling strength $J$ can be about 1 meV in semiconductor quantum dots[20] and 6 meV in molecules[21]. Therefore, with semiconductor quantum dot implementation, $T = 1.1$ K if we want $p_i = 10^{-9}$, and $T = 6.5$ K if $p_i = 0.03$, which is the maximum error probability that may be handled with the most sophisticated error correction schemes available today[17]. Conversely, if we operate at 1.1 K, then $p_i$ can be as low as $10^{-9}$ in semiconductor quantum dot based systems. Room temperature operation will require $J = 270$ meV with $p_i = 10^{-9}$ and $J = 46$ meV with $p_i = 0.03$. Neither value of $J$ is achievable with semiconductor quantum dots or molecular magnets at present, which unfortunately precludes room temperature operation with present day technology. Future technological advances may make room temperature operation feasible.

If we operate at 1.1 K with $p_i = 10^{-9}$, then $Z = g\mu_B B_{global} = kTln(1/p_i) = 1$ meV. Here, $g$ is the g-factor of the quantum dot material and $B_{global}$ is the magnetic flux density of the global magnetic field. We can make $B_{global}$ small by using materials with large g-factors. If we use $InSb_{1-x}N_x$, as the quantum dot material, which is predicted to have a g-factor of 900 [22], then $B_{global} = 0.04$ Tesla if $Z = 1$ meV.

The local magnetic fields needed to write input bits in dots A and C need to be approximately 10 times larger than the global field (see Appendix). Therefore, the local field strengths need not exceed 0.4 Tesla, which should be within reach of magnetic random access memory technology[23].

Finally, one concern is that using a material with giant g-factor may adversely affect the spin flip time. But it will affect $T_1$ and $T_1^*$ almost equally. Therefore, the extrinsic error probability $p_e = T_1^*/T_1$, will not change by much. If $T_1^*$ goes down, then the maximum clock frequency $1/T_1^*$ will increase commensurately.

## 7. Conclusion

We have shown that a simple linear array of 3 spins in quantum dots, with nearest neighbor exchange coupling, realizes the universal classical Boolean NAND gate, if placed in a global dc magnetic field and allowed to relax to the thermodynamic ground state. Recent advances in single spin electronics, allowing control over single electrons[24-26], has brought us close to the realization of such computing elements. The energy dissipated during switching between states is $kTln(1/p_i)$. With $p_i$ as low as $10^{-9}$, this energy is ~21 $kT$ which is much better than the 40,000 – 50,000 $kT$ dissipated in present day transistor based gates[27]. A serious drawback however is that the temperature of operation is ~ 1 K due to present constraints in quantum dot technology. At this temperature, the energy dissipated during switching is ~ $3\times10^{-22}$ Joules if $p_i = 10^{-9}$. This can extend Moore's law easily into the next few decades. We also point out that realization of these gates does not require phase coherence of spin, which is difficult to preserve over long times, even at low temperatures. This paradigm is completely classical unlike quantum computing; therefore, these gates are considerably easier to implement than quantum gates.

**Acknowledgement:** This work is supported by the US Air Force Office of Scientific Research under grant FA9550-04-1-0261 and by the US National Science Foundation under grant CCF-0726373. The conclusions in this paper are of the authors and do not represent an endorsement by the National Science Foundation.

**References**

1. Moore G. E. 1965 *Electronics Magazine* (McGraw Hill, New York) April 19.
2. Bandyopadhyay S, Das B and Miller A E 1994 *Nanotechnology* **5** 113.
3. Keyes R W and Landauer R 1970 *IBM J. Res. Develop.* **14** 152.
4. Rugar D, Budakian R, Mamin H J and Chui B H 2004 *Nature* **430** 329.
5. Elzerman J M et al. 2004 *Nature* **430** 431.
6. Xiao M, Martin I, Yablonovitch E and Jiang H W 2004 *Nature* **430** 435.
7. Molotkov S N and Nazin S S 1997 *Phys. Low Dim. Struct.* **10** 85.
8. Molotkov S N and Nazin S S 1995 *JETP Lett.* **62**, 273.
9. Rafailov E U et al. 2006 Appl. Phys. Lett., **88**, 041101; Markus A et al, 2003 Appl. Phys. Lett., **82**, 1818.
10. Amasha S. et al. http://www.arxiv.org/abs/cond-mat/0607110 and http://xxx.lanl.gov/abs/0707.1656v1.
11. Bakshi P, Broido D and Kempa K 1991 *J. Appl. Phys*. **70** 5150. The logic gate idea with quantum dashes was communicated by P. Bakshi in a private communication.
12. Bandyopadhyay S and Roychowdhury V P 1997 *Superlat. Microstruct.* **22**, 411.
13. Bandyopadhyay S 2005 *Superlat. Microstruct.*, **37** 77.
14. Anantram M and Roychowdhury V P 1999 *J. Appl. Phys*. **85** 1622.
15. Pramanik S, et al. 2007 *Nature Nanotech*. **2**, 216.
16. deSousa R and Das Sarma S 2003 Phys. Rev. B **67** 033301.
17. Knill E 2005 *Nature* **434** 39.
18. Salahuddin S and Datta S 2007 *Appl. Phys. Lett.* **90** 093503.
19. Cavin R K, Zhirnov V V, Hutchby J A and Bourianoff G I, 2005 Fluctuations and Noise Lett. **5** C29.
20. Melnikov D V and Leburton J-P 2006 *Phys. Rev. B* **73** 155301.
21. Hirjibehedin C F, Lutz C P and Heinrich A J 2006 *Science* **312** 1021.

22. Zhang X W, Fan W J, Li S S and Xia J B 2007, *Appl. Phys. Lett.* **90** 193111.

23. Nembach H T, et al. 2005 *Appl. Phys. Lett.* **87** 142503.

24. Petta J R et al. 2005 *Science* **309** 2180.

25. Koppens F H L et al. 2006 *Nature* **442** 766.

26. Hanson R, Witkamp B, Vandersypen L M K, van Beveren L H W, Elzerman J M and Kouwenhoven L P, 2003 *Phys. Rev. Lett.* **91** 196802.

27. The International Technology Roadmap for Semiconductors published by the Semiconductor Industry Association. http://www.itrs.net.

**APPENDIX**

We tabulate below the 8 many-body eigenenergies ($E_n$) and the eigenstates ($\psi_n$) of the 3-spin array for the four cases corresponding to the four input combinations shown in Figs. 1(a) – 1(d).

**Case I:** $h_A = h_C = h>0$: This is the case when the input bits are [1 1] and the situation corresponds to Fig. 1(a) [the first entry in the truth table of the NAND gate]. The 8 eigenenergies $E_n$ and eigenstates $\psi_n$ are:

**Table A1: Eigenenergies and eigenstates when the inputs are [1 1] ; $h_A = h_C = h>0$.**

| Eigenenergies ($E_n$) | Eigenstates ($\psi_n$) |
|---|---|
| $-J - h - Z - \Delta_1$ | $[0, 2/\beta_1, \alpha_1/(J\beta_1), 0, 2/\beta_1, 0, 0, 0]$ |
| $2J - 2h - 3Z$ | $[1, 0, 0, 0, 0, 0, 0, 0]$ |
| $-J + h + Z - \Delta_2$ | $[0, 0, 0, 2/\beta_3, 0, -\alpha_3/(J\beta_3), 2/\beta_3, 0]$ |
| $-Z$ | $[0, 1/\sqrt{2}, 0, 0, -1/\sqrt{2}, 0, 0, 0]$ |
| $Z$ | $[0, 0, 0, -1/\sqrt{2}, 0, 0, 1/\sqrt{2}, 0]$ |
| $-J - h - Z + \Delta_1$ | $[0, 2/\beta_6, \alpha_6/(J\beta_6), 0, 2/\beta_6, 0, 0, 0]$ |
| $-J + h + Z + \Delta_2$ | $[0, 0, 0, 2/\beta_7, 0, -\alpha_7/(J\beta_7), 2/\beta_7, 0]$ |
| $2J + 2h + 3Z$ | $[0, 0, 0, 0, 0, 0, 0, 1]$ |

where

$$\Delta_1 = \sqrt{(h+J)^2 + 8J^2}$$

$$\Delta_2 = \sqrt{(h-J)^2 + 8J^2}$$

$$\alpha_1 = -J - h - \Delta_1$$

$$\alpha_3 = J - h + \Delta_2$$

$$\alpha_6 = -J - h + \Delta_1$$

$$\alpha_7 = J - h - \Delta_2$$

$$\beta_n = \sqrt{(\alpha_n/J)^2 + 8}$$

Note that the eigenenergies $E_n$ depend on $Z$, but the eigenstates $\psi_n$ do not. In Table A1, the eigenenergies are arranged in ascending order (i.e. the first entry is the ground state and the last entry is the highest excited state), *provided $h \gg J$ and $J > Z/2$*. The last inequality ensures that

$$2J - 2h - 3Z > -J - h - Z - \Delta_1 \ ,$$

which guarantees that the first excited state is higher in energy than the ground state. We now address why we need $h \gg J$.

Note that the ground state wavefunction is the entangled state

$$\psi_{ground}^{11} = \frac{2}{\beta_1}|\downarrow\downarrow\uparrow\rangle + \frac{\alpha_1}{J\beta_1}|\downarrow\uparrow\downarrow\rangle + \frac{2}{\beta_1}|\uparrow\downarrow\downarrow\rangle$$

However, when the inputs are [1, 1], or [$\downarrow,\downarrow$], we want the output to be [0], or [$\uparrow$] since this is the situation shown in Fig. 1(a). Therefore, the *desired* ground state is the unentangled state

$$\psi_{desired}^{11} = |\downarrow\uparrow\downarrow\rangle = |\downarrow\rangle \otimes |\uparrow\rangle \otimes |\downarrow\rangle$$

Obviously, we can make $\psi_{ground}^{11} \approx \psi_{desired}^{11}$ if

$$\left|\frac{\alpha_1}{2J}\right| = \frac{h + J + \sqrt{(h+J)^2 + 8J^2}}{2J} \gg 1$$

i.e. if $h \gg J$. In other words, the 3-spin configuration in Fig. 1(a) will become the ground state if we make $h \gg J$. In that case, whenever we apply the inputs [1, 1] to the input dots **A** and **C** and let the system relax thermodynamically to the ground state (by emitting phonons, etc.), it will reach the state in Fig. 1(a) where the output bit (in dot B) will be [0] and *we will have realized the first entry in the truth table of the NAND gate.*

**Case II: $h_A = h_C = -h < 0$:** This is the case when the input bits are [0 0] and the situation corresponds to Fig. 1(b). In this case, the eigenenergies and eigenstates are obtained by replacing the quantity $h$ in Table S1 with $-h$.

Table A2: Eigenenergies and eigenstates when the inputs are [0 0]; $h_A = h_C = -h < 0$

| Eigenenergies | Eigenstates |
| --- | --- |
| $-J - h + Z - \Delta_1$ | $[0, 0, 0, 2/\beta_1, 0, \alpha_1/(J\beta_1), 2/\beta_1, 0]$ |
| $2J - 2h + 3Z$ | $[0, 0, 0, 0, 0, 0, 0, 1]$ |
| $-J + h - Z - \Delta_2$ | $[0, 2/\beta_3, \alpha_3/(J\beta_3), 0, 2/\beta_3, 0, 0, 0]$ |
| $-Z$ | $[0, -1/\sqrt{2}, 0, 0, 1/\sqrt{2}, 0, 0, 0]$ |
| $Z$ | $[0, 0, 0, -1/\sqrt{2}, 0, 0, 1/\sqrt{2}, 0]$ |
| $-J - h + Z + \Delta_1$ | $[0, 0, 0, 2/\beta_6, 0, \alpha_6/(J\beta_6), 2/\beta_6, 0]$ |
| $-J + h - Z + \Delta_2$ | $[0, 2/\beta_7, \alpha_7/(J\beta_7), 0, 2/\beta_7, 0, 0, 0]$ |
| $2J + 2h - 3Z$ | $[1, 0, 0, 0, 0, 0, 0, 0]$ |

For this case, the ground state wavefunction is the entangled state

$$\psi_{ground}^{00} = \frac{2}{\beta_1}\left|\downarrow\uparrow\uparrow\right\rangle + \frac{\alpha_1}{J\beta_1}\left|\uparrow\downarrow\uparrow\right\rangle + \frac{2}{\beta_1}\left|\uparrow\uparrow\downarrow\right\rangle$$

whereas the desired state shown in Fig. 1(b) is the unentangled state

$$\psi_{desired}^{00} = |\uparrow\downarrow\uparrow\rangle = |\uparrow\rangle \otimes |\downarrow\rangle \otimes |\uparrow\rangle$$

Once again, we can make $\psi_{ground}^{00} \approx \psi_{desired}^{00}$ if we make $|\alpha_1/2J| \gg 1$, or $h \gg J$. Then, if we apply inputs [0 0] to dots **A** and **C**, and let the system relax to the ground state, dot **B** will have output [1] corresponding to Fig. 1(b), and we will have realized the second entry in the truth table of the NAND gate. All we need for this to happen is $h \gg J$.

**Case III: $-h_A = h_C = h > 0$:** This is the case when the input bits are [0 1] and the situation corresponds to Fig. 1(c). In this case, the eigenenergies and eigenstates are more complicated and given in Table A3.

Table A3: Eigenenergies and eigenstates when the inputs are [0 1]; $-h_A = h_C = h > 0$

| Eigenenergies | Eigenstates |
|---|---|
| $-\theta_4 - 2J/3 - Z + \left(\sqrt{3}i/2\right)\theta_3$ | $[0, \pi_3^{(1)}/(J^2\pi_4^{(1)}), 2\pi_1^{(1)}/(J\pi_4^{(1)}), 0, 4/\pi_4^{(1)}, 0, 0, 0]$ |
| $-\theta_4 - 2J/3 + Z + \left(\sqrt{3}i/2\right)\theta_3$ | $[0, 0, 0, \pi_3^{(2)}/(J^2\pi_4^{(2)}), 0, 2\pi_1^{(2)}/(J\pi_4^{(2)}), 4/\pi_4^{(2)}, 0]$ |
| $-\theta_4 - 2J/3 - Z - \left(\sqrt{3}i/2\right)\theta_3$ | $[0, \pi_3^{(3)}/(J^2\pi_4^{(3)}), 2\pi_1^{(3)}/(J\pi_4^{(3)}), 0, 4/\pi_4^{(3)}, 0, 0, 0]$ |
| $-\theta_4 - 2J/3 + Z - \left(\sqrt{3}i/2\right)\theta_3$ | $[0, 0, 0, \pi_3^{(4)}/(J^2\pi_4^{(4)}), 0, 2\pi_1^{(4)}/(J\pi_4^{(4)}), 4/\pi_4^{(4)}, 0]$ |
| $2J - 3Z$ | $[1, 0, 0, 0, 0, 0, 0, 0]$ |
| $2J + 3Z$ | $[0, 0, 0, 0, 0, 0, 0, 1]$ |
| $\theta_4 - 2J/3 - Z$ | $[0, \pi_3^{(7)}/(J^2\pi_4^{(7)}), 2\pi_1^{(7)}/(J\pi_4^{(7)}), 0, 4/\pi_4^{(7)}, 0, 0, 0]$ |
| $\theta_4 - 2J/3 + Z$ | $[0, 0, 0, \pi_3^{(8)}/(J^2\pi_4^{(8)}), 0, 2\pi_1^{(8)}/(J\pi_4^{(8)}), 4/\pi_4^{(8)}, 0]$ |

where,

$$\theta_1 = J\left[9(h/J)^2 - 10 + 3i\sqrt{3(h/J)^6 + 12(h/J)^4 + 69(h/J)^2 + 27}\right]^{1/3}$$

$$\theta_2 = -\frac{4J^2}{3\theta_1}\left[(h/J)^2 + 7/3\right]$$

$$\theta_3 = \frac{2\theta_1}{3} + \frac{3\theta_2}{2} = 2i\,\mathrm{Im}\left(\frac{2\theta_1}{3}\right)$$

$$\theta_4 = \frac{2\theta_1}{3} - \frac{3\theta_2}{2} = 2\,\mathrm{Re}\left(\frac{2\theta_1}{3}\right)$$

$$\pi_1^{(1)} = -\theta_4/2 - 2J/3 + 2h + (\sqrt{3}i/2)\theta_3$$
$$\pi_2^{(1)} = -\theta_4/2 - 2J/3 - Z + (\sqrt{3}i/2)\theta_3$$
$$\pi_3^{(1)} = \left[\pi_2^{(1)}\right]^2 + 2\pi_2^{(1)}(Z + J + h) + 4Jh + 2JZ - 4J^2 + Z^2 + 2hZ$$

$$\pi_1^{(2)} = \pi_1^{(1)}$$
$$\pi_2^{(2)} = \pi_2^{(1)} + 2Z$$
$$\pi_3^{(2)} = \left[\pi_2^{(2)}\right]^2 + 2\pi_2^{(2)}(-Z + J + h) + 4Jh - 2JZ - 4J^2 + Z^2 - 2hZ$$

$$\pi_1^{(3)} = \pi_1^{(1)} - \sqrt{3}i\theta_3$$
$$\pi_2^{(3)} = \pi_2^{(1)} - \sqrt{3}i\theta_3$$
$$\pi_3^{(3)} = \left[\pi_2^{(3)}\right]^2 + 2\pi_2^{(3)}(Z + J + h) + 4Jh + 2JZ - 4J^2 + Z^2 + 2hZ$$

$$\pi_1^{(4)} = \pi_1^{(2)} - \sqrt{3}i\theta_3$$
$$\pi_2^{(4)} = \pi_2^{(2)} - \sqrt{3}i\theta_3$$
$$\pi_3^{(4)} = \left[\pi_2^{(4)}\right]^2 + 2\pi_2^{(4)}(-Z + J + h) + 4Jh - 2JZ - 4J^2 + Z^2 - 2hZ$$

$$\pi_1^{(7)} = \theta_4 - 2J/3 + 2h$$
$$\pi_2^{(7)} = \theta_4 - 2J/3 - Z$$
$$\pi_3^{(7)} = \left[\pi_2^{(7)}\right]^2 + 2\pi_2^{(7)}(Z + J + h) + 4Jh + 2JZ - 4J^2 + Z^2 + 2hZ$$

$$\pi_1^{(8)} = \pi_1^{(7)}$$
$$\pi_2^{(8)} = \pi_2^{(7)} + 2Z$$
$$\pi_3^{(8)} = \left[\pi_2^{(8)}\right]^2 + 2\pi_2^{(8)}(-Z + J + h) + 4Jh - 2JZ - 4J^2 + Z^2 - 2hZ$$

$$\pi_4^{(n)} = \left[ \frac{\left[\pi_3^{(n)}\right]^2}{J^4} + \frac{4\left[\pi_1^{(n)}\right]^2}{J^2} + 16 \right]^{\frac{1}{2}} \quad (n = 1\ldots\ldots 8).$$

The ground state wavefunction is given by the entangled state

$$\psi_{ground}^{01} = \frac{\pi_3^{(1)}}{J^2 \pi_4^{(1)}} |\downarrow\downarrow\uparrow\rangle + \frac{2\pi_1^{(1)}}{J\pi_4^{(1)}} |\downarrow\uparrow\downarrow\rangle + \frac{4}{\pi_4^{(1)}} |\uparrow\downarrow\downarrow\rangle$$

whereas the desired state shown in Fig. 1(c) is the unentangled state

$$\psi_{desired}^{01} = |\uparrow\downarrow\downarrow\rangle = |\uparrow\rangle \otimes |\downarrow\rangle \otimes |\downarrow\rangle$$

Once again, we can make $\psi_{ground}^{01} \approx \psi_{desired}^{01}$ if we make $h >> J$. Therefore, if we apply inputs [0 1] to dots **A** and **C** and let the system relax to the ground state, dot **B** will have output [1] corresponding to Fig. 1(c), and we will have realized the third entry in the truth table of the NAND gate.

**Case IV: - $h_A = h_C = - h<0$:** This is the case when the input bits are [1 0] and the situation corresponds to Fig. 1(d). The eigenenergies do not change from Table A3 since they depend on $h^2$ and are therefore insensitive to the sign of $h$. However, the eigenstates are sensitive to the sign of $h$ and change. The eight eigenstates can be found by replacing $\pi_p^{(q)}$ by $\hat{\pi}_p^{(q)}$ where,

$$\hat{\pi}_p^{(q)}(h) = \pi_p^{(q)}(-h) \quad (p=1\ldots 4,\ q=1\ldots 8).$$

The ground state wavefunction is given by the entangled state

$$\psi_{ground}^{10} = \frac{\hat{\pi}_3^{(1)}}{J^2 \hat{\pi}_4^{(1)}} |\downarrow\downarrow\uparrow\rangle + \frac{2\hat{\pi}_1^{(1)}}{J\hat{\pi}_4^{(1)}} |\downarrow\uparrow\downarrow\rangle + \frac{4}{\hat{\pi}_4^{(1)}} |\uparrow\downarrow\downarrow\rangle$$

while the desired state shown in Fig. 1(d) is the unentangled state

$$\psi_{desired}^{10} = |\downarrow\downarrow\uparrow\rangle = |\downarrow\rangle \otimes |\downarrow\rangle \otimes |\uparrow\rangle$$

It is easy to check that we can make $\psi_{ground}^{10} \approx \psi_{desired}^{10}$ if we make $h \gg J$. Therefore, if we apply inputs [1 0] to dots **A** and **C** and let the system relax to the ground state, dot **B** will have output [1] corresponding to Fig. 1(d), and we will have realized the fourth and final entry in the truth table of the NAND gate.

In conclusion, what we have shown here is that if we place a 3-spin array with nearest neighbor exchange coupling in a dc magnetic field, and align the spins in the two peripheral dots (designated as input ports) with sufficiently strong local magnetic field $B_{local}$ such that the Zeeman splitting in the inputs dots $2h = g\mu_B B_{local}$ is much larger than $2J$, then the spin polarization in the output (central) dot will always conform to the NAND function of the two inputs, once the array relaxes to the thermodynamic ground state. This realizes a "single-spin-NAND-gate".

One final issue that needs to be resolved is the following. In order for the NAND gate to work correctly, we need that $h \gg J$. How large should the ratio $h/J$ be? Note from Table A1 that the ground state approaches the unentangled state $|\downarrow\uparrow\downarrow\rangle$ if $|\alpha_1/(J\beta_1)| \to 1$ and $|2/\beta_1| \to 0.$ Let us define $P_{\downarrow\uparrow\downarrow} = |\alpha_1/(J\beta_1)|^2$ and $P_{\downarrow\downarrow\uparrow} = P_{\uparrow\downarrow\downarrow} = |2/\beta_1|^2$ since $\alpha_1/(J\beta_1)$ is the amplitude of the $|\downarrow\uparrow\downarrow\rangle$ component and $2/\beta_1$ is the amplitude of the $|\downarrow\downarrow\uparrow\rangle$ or $|\uparrow\downarrow\downarrow\rangle$ components in the ground state. In Fig. A1, we plot these quantities as a function of the ratio $h/J$. Note that $|\alpha_1/(J\beta_1)| \to 1$ and $|2/\beta_1| \to 0$ when $h/J \geq 10$. Therefore, in Case I, we need $h/J \geq 10$ to make the ground state nearly indistinguishable from the unentangled state $|\downarrow\uparrow\downarrow\rangle$.

In Figs. A2, A3 and A4, we plot the equivalent quantities for Cases II, III and IV, respectively. Once again, we find that ensuring $h/J \geq 10$ is sufficient.

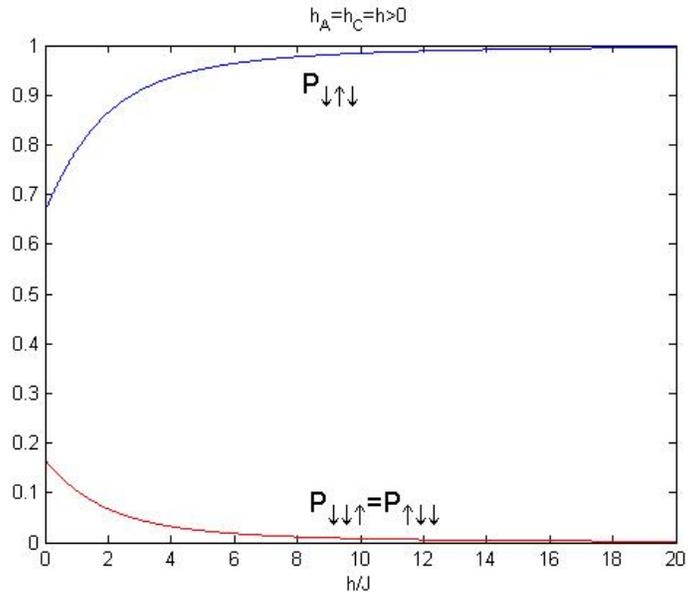

Fig A1: Probabilities as a function of the ratio *h/J* when the inputs bits are [1 1]

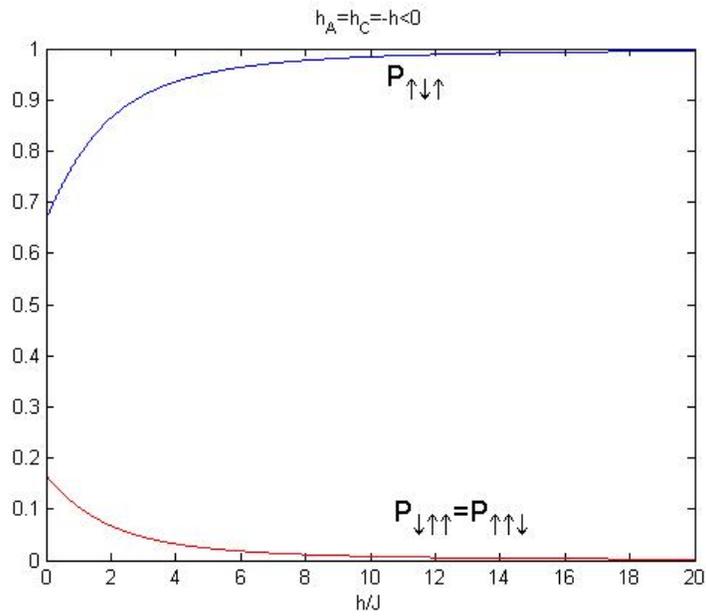

Fig A2: Probabilities as a function of the ratio *h/J* when the inputs bits are [0 0]
23

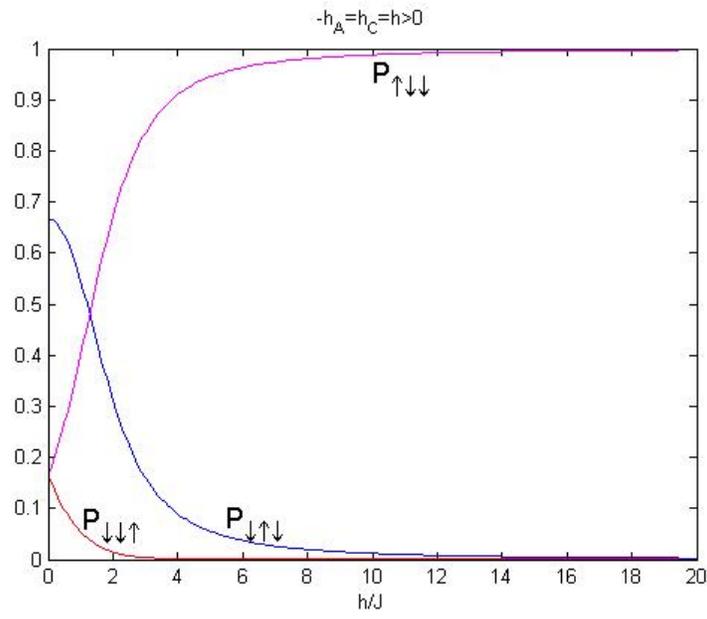

Fig A3: Probabilities as a function of the ratio *h/J* when the inputs bits are [0 1]

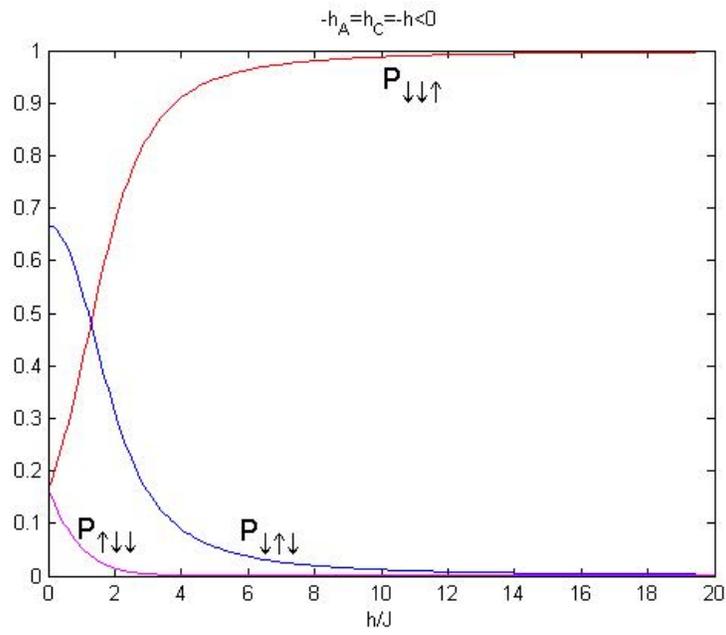

Fig A4: Probabilities as a function of the ratio *h/J* when the inputs bits are [1 0]

Therefore, in all cases, it is adequate to have the strengths of the local magnetic fields writing inputs bits no more than 10 times stronger than the global magnetic field. If the global magnetic flux density is 0.04 Tesla, it is sufficient to have the local magnetic flux density 0.4 Tesla.